\documentclass[final]{aipproc}
\usepackage{epsfig}
\layoutstyle{6x9}
\SetInternalRegister\hbadness{8000}
\begin{document}

\title[Strangeness content]{Determination of the strangeness content of 
light-flavour isoscalars from their production rates in hadronic Z decays 
at LEP}


\author{Vladimir~A.~Uvarov}{
  address={Institute for High Energy Physics, RU-142284, Protvino, Russia},
  email={uvarov@mx.ihep.su}
}

\copyrightyear {2001}

\begin{abstract}
A new phenomenological approach is suggested for determining the strangeness 
content of light-flavour isoscalars.
This approach is based on phenomenological laws of hadron production related 
to the spin, isospin, strangeness content and mass of the particles.
From the total production rates per hadronic Z decay of all light-flavour 
hadrons, measured so far at LEP, the values of the nonstrange-strange mixing 
angles are found to be
$\vert\varphi_P\vert$ = 42.3$^{\circ}$$\pm$3.5$^{\circ}$,
$\vert\varphi_V\vert$ = 10$^{\circ}$$\pm$8$^{\circ}$,
$\vert\varphi_T\vert$ = 16$^{\circ}$$\pm$11$^{\circ}$ and
$\vert\varphi_S\vert$ = 13$^{\circ}$$\pm$9$^{\circ}$.
Our results on the $\eta$--$\eta^{\,\prime}$, $\omega$--$\phi$ and 
$f_2^{~}$--$f_2^{\,\prime}$ isoscalar mixing are consistent with the present
experimental evidence. 
The strangeness content obtained for the $f_0$(980) scalar/isoscalar is not 
consistent with the values supported by recent model studies and is 
discussed further in the framework of our approach and the $K$-matrix analysis.
\end{abstract}

\date{\today}
\maketitle

The quark contents of pseudoscalar ($P$), vector ($V$), tensor ($T$)
and scalar ($S$) mesons have been discussed many times since the discovery 
of unitary SU(3)-flavour symmetry. 
This is still quite an interesting question because the quark contents of the 
lightest isoscalars differ from the predictions of the SU(3) quark 
model and are very important SU(3)-breaking hadronic parameters.

In terms of the $n\bar{n} = (u\bar{u}+d\bar{d})/\sqrt{2}$ and $s\bar{s}$
quark basis, the strangeness contents of the physical isoscalars are given by 
the {\it nonstrange-strange mixing angles} $\varphi_P$, $\varphi_V$, 
$\varphi_T$ and $\varphi_S$.
Assuming orthogonality of the isoscalar partners and no mixing with other 
states and glueballs, the flavour wave functions of the 
$\eta$--$\eta^{\,\prime}$ pseudoscalars are defined to be
$\eta = n\bar{n}\cdot\cos\varphi_P - s\bar{s}\cdot\sin\varphi_P$ and  
$\eta^{\,\prime} = n\bar{n}\cdot\sin\varphi_P + s\bar{s}\cdot\cos\varphi_P$.
The flavour wave functions of the $\omega$--$\phi$ vectors 
($f_2^{~}$--$f_2^{\,\prime}$ tensors) are defined in a way analogous to the 
$\eta$--$\eta^{\,\prime}$ case, replacing $\eta \rightarrow \omega (f_2^{~})$, 
$\eta^{\,\prime} \rightarrow \phi (f_2^{\,\prime})$ and
$\varphi_P \rightarrow \varphi_V (\varphi_T)$. 
The flavour wave function of the $f_0(980)$ scalar is written as
$f_0(980) = n\bar{n}\cdot\cos\varphi_S + s\bar{s}\cdot\sin\varphi_S$.

The values of these mixing angles have been estimated from different  
phenomenological and theoretical analyses (see Refs. [1-13] and references 
therein).
Estimates of the mixing angle $\varphi_P$ were obtained from the 
available world data on different decay processes.
The corresponding average value was found to be 
$\varphi_P$ = 39.2$^{\circ}$$\pm$1.3$^{\circ}$ \cite{cp2,cp3}.
For the mixing angles $\varphi_V$ and $\varphi_T$, most theoretical 
and phenomenological analyses (see, e.g., Refs. [4\,-\,6,\,13]) 
predict values which are close to the ``ideal'' mixing: 
$\varphi_V \simeq +3.4^{\circ}$ and $\varphi_T \simeq -7.3^{\circ}$.
The interpretation of the $f_0$(980) scalar is one of the most controversial 
ones in meson spectroscopy \cite{cp7}. 
The question is whether the $f_0$(980) consists mostly of 
$n\bar{n}$ or of $s\bar{s}$ states.
The phenomenological analyses of the experimental data on different decay 
processes [9\,-12] favour the $s\bar{s}$ dominance of the $f_0$(980). 

Recently the strangeness contents of light-flavour isoscalars have been 
obtained for the first time \cite{cp14} from their total production rates per 
hadronic Z decay at LEP.
Here only some key points and the final results of that analysis \cite{cp14}
are discussed.
 
It has been shown [15\,-19] that the total production rates per hadronic Z 
decay ($\langle n \rangle$) of all light-flavour mesons ($M$) and baryons 
($B$), measured so far at LEP, follow phenomenological laws related 
to the spin ($J$), isospin ($I$), strangeness content ($k$) and mass ($m$) of 
the particles. These regularities can be combined into one empirical formula:
\begin{equation}
\label{neq3}
{\langle n \rangle} \,=\, 
A\cdot\beta_H\cdot(2J+1)\cdot\gamma^{\,k}\cdot\exp{[-b_H(m/m_0)^{N_H}]},
\end{equation}
where $H$ = $M$ or $B$, $m_0$ = 1 GeV, $\gamma$ is the strangeness 
suppression factor with a value of $\gamma \simeq 0.5$ for all hadrons,
$k$ is the number of $s$ and $\bar{s}$ quarks in the hadron, and $b_H$
is the slope of the $m$ dependence.
The values of the power $N_H$ and of the coefficient $\beta_H$ are different 
for mesons and baryons: $N_M$ = 1, $N_B$ = 2 and $\beta_M$ = 1, 
$\beta_B$ = $4/(C_{\pi/p}\lambda_{QS})$, 
where $\lambda_{QS}$ = $(2J+1)(2I+1)$ can be interpreted as a fermion 
suppression factor originating from the quantum statistics properties of bosons 
and fermions, and $C_{\pi/p}$ is the $\pi$/p ratio at the zero mass limit 
with a value of $C_{\pi/p} \simeq 3$ which could be expected from quark
combinatorics. 
The slope $b_M$ has two values: one for vector, tensor and scalar mesons and 
the other for pseudoscalar mesons. 
In spite of this ``splitting'', Eq.~(\ref{neq3}) assumes the validity of the 
relation $V/P = 3$ at the zero mass limit.
This difference in slopes can probably be explained by the influence of the 
spin-spin interaction between the quarks of the meson. 
However, there is no influence of the spin-orbital interaction of the quarks. 

The {\it purpose of this analysis} \cite{cp14} is to determine the mixing 
angles of the $\eta$--$\eta^{\,\prime}$, $\omega$--$\phi$, 
$f_2^{~}$--$f_2^{\,\prime}$ and $f_0$(980) isoscalars from the 
{\it simultaneous} fit of Eq.~(\ref{neq3}) to the total production rates per 
hadronic Z decay of all light-flavour hadrons measured so far at LEP, assuming 
that the mixing angle $\varphi$ and the strangeness contents $k_1$ and $k_2$ 
of the isoscalar partners are related by: 
$k_1 = 2\sin^2\varphi$ and $k_2 = 2\cos^2\varphi$.
The strangeness contents of baryons and $I$$\neq$0 mesons are taken from the 
predictions of the SU(3) quark model.

The total production rates {\it per isospin state} $\langle n \rangle$, used 
in the fit, were obtained by averaging the total production rates $\bar{n}$ of 
hadrons belonging to the same isomultiplet. 
The rates $\bar{n}$ themselves were obtained for at 
least one state of a given isomultiplet as a weighted-average of the 
measurements of the four LEP experiments (see details in Ref.~\cite{cp14} and 
references therein). 

Firstly we tested the sensitivity of Eq.~(\ref{neq3}) to the values of the 
power $N_H$. 
In the test fit the following parameters were fixed:
$\varphi_P$ = 44.7$^{\circ}$, $\varphi_V$ = 3.7$^{\circ}$, 
$\varphi_T$ = $-7.3^{\circ}$
predicted by the quadratic Gell-Mann--Okubo mass formula \cite{cp13},
$\varphi_S$ = 0 suggested as an ad hoc value in \cite{U1}, and
$C_{\pi/p}$ = 3 predicted by quark combinatorics for production of direct or 
massless particles.
The values obtained, $N_M$ = 1.04$\pm$0.06 and $N_B$ = 2.07$\pm$0.05, strongly 
suggest the use of the fixed values of $N_M$ = 1 and $N_B$ = 2 in our approach.

\begin{figure}[t]
\centering\mbox{\epsfig{file=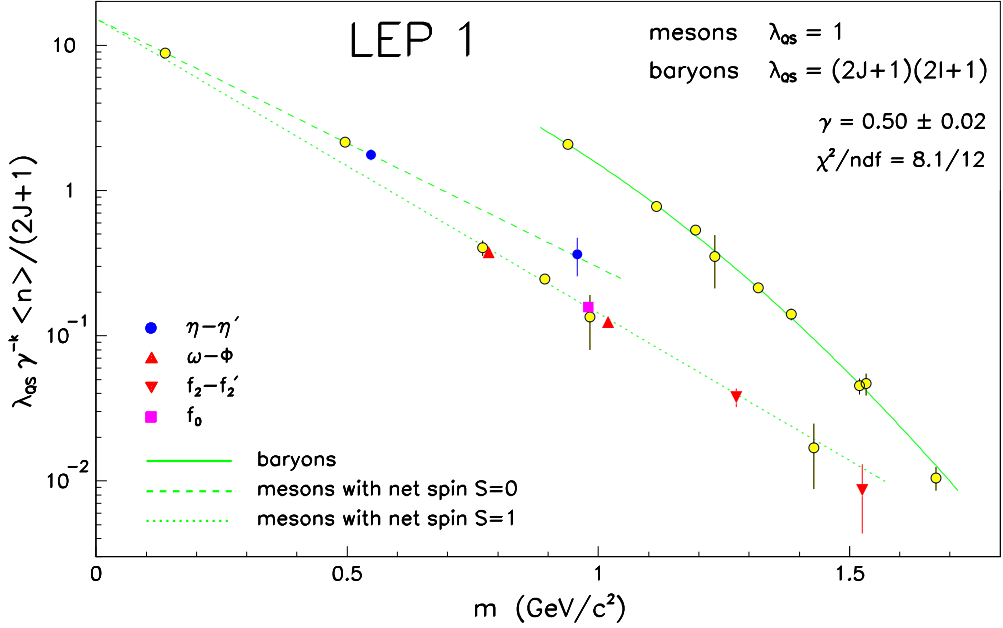,width=0.9\textwidth}}
\caption{Mass dependences of the total production rates per spin and isospin 
state for hadrons in hadronic Z decays weighted by a factor 
$\lambda_{QS} \gamma^{-k}$, where $\lambda_{QS}=1$ for mesons and  
$\lambda_{QS}=(2J+1)(2I+1)$ for baryons. The references for all data points 
can be found in Ref.~\cite{cp14}.}
\label{fig1}
\end{figure}
In the final fit only the values of $N_M$ = 1 and $N_B$ = 2 are fixed, all 
other parameters are free.
This fit (with $\chi^2$/dof = 8.1/12) is illustrated in Fig.~\ref{fig1},
where the three curves are the result of the fit for baryons, for mesons with 
net spin S\,=\,0 (pseudoscalars), and for mesons with net spin S\,=\,1 
(vectors, tensors and scalars).
The values of the strangeness suppression factor and of the $\pi$/p ratio at 
the zero mass limit are found to be $\gamma$ = 0.50$\pm$0.02 and 
$C_{\pi/p}$ = 2.8$\pm$0.2, in good agreement with our previous results [17-19].
The values of the isoscalar mixing angles are found to be
$\vert\varphi_P\vert$ = 42.3$^{\circ}$$\pm$3.5$^{\circ}$,
$\vert\varphi_V\vert$ = 10$^{\circ}$$\pm$8$^{\circ}$,
$\vert\varphi_T\vert$ = 16$^{\circ}$$\pm$11$^{\circ}$ and
$\vert\varphi_S\vert$ = 13$^{\circ}$$\pm$9$^{\circ}$.
Quite remarkably, our values of the $\eta$--$\eta^{\,\prime}$, 
$\omega$--$\phi$ and $f_2^{~}$--$f_2^{\,\prime}$ mixing angles are compatible 
within the errors with the predictions of theoretical and phenomenological 
analyses (see, e.g., Refs. [2\,-\,6,\,13]).
However, our value of the mixing angle $\varphi_S$ is not consistent 
with the results of recent phenomenological studies which favour the 
$s\bar{s}$ dominance of the $f_0$(980) scalar [9\,-12]. 

The disagreement obtained for the $f_0$(980) can probably be related to a 
question which is still open. 
This is whether the $f_0$(980) belongs to the scalar $q\bar{q}$ nonet 
$1^3P_0$ or whether it should be considered as an exotic state.
The scalar nonet classification can be performed in terms of so-called
``bare states'' (the $K$-matrix poles) \cite{cp8,K1}. 
In this way the scalar/isoscalars have been found to be 
$f_0^{bare}$(720$\pm$100) and $f_0^{bare}$(1260$\pm$30) with the mixing angle 
$\varphi_S[f_0^{bare}(720)]$ = ${-70^{\circ}}_{-16^{\circ}}^{+~5^{\circ}}$.

\begin{figure}[t]
\centering\mbox{\epsfig{file=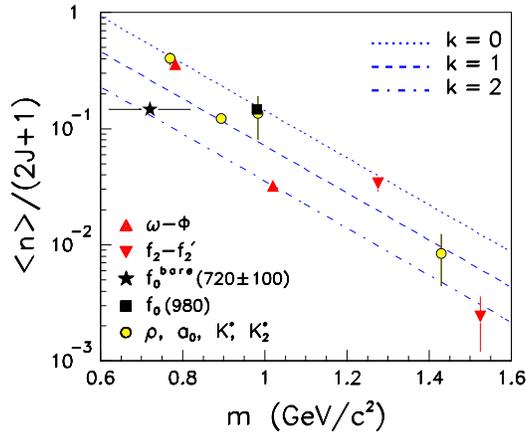,width=0.47\textwidth}}
\caption{Total production rate per spin and isospin state as a function 
of $m$ for vector, tensor and scalar mesons in hadronic Z decays. Curves 
are the result of the final fit with $k$ = 0, $k$ = 1 and $k$ = 2.} 
\label{fig2}
\end{figure}

As was observed by Montanet \cite{G70}, 
the $f_0$(980) data point (Fig.~\ref{fig2}) shifted from the real mass to 
the ``bare mass'' (720 MeV) is close to the line corresponding to the mesons 
with $k$ = 2.
This interesting relation between our phenomenology and the $K$-matrix 
analysis can probably be considered as a {\it speculative} argument 
for re-determining the scalar mixing angle, replacing in the fit the real 
mass of the $f_0$(980) by the ``bare'' one. 
Such a fit gives a scalar mixing angle of
$\vert\varphi_S^{\,bare}\vert$ = 73$^{\circ}$$\pm$7$^{\circ}$$\pm$24$^{\circ}$,
where the second error is due to the uncertainty ($\pm$100 MeV) of the 
``bare mass''. This value is well consistent with the predictions of  
recent phenomenological studies [8,\,10\,-12].

In conclusion, a new phenomenological approach has been suggested for 
determining the mixing angles of light-flavour isoscalars.
For the first time, the total production rates per hadronic Z decay of all 
light-flavour hadrons, measured so far at LEP, have been used for this purpose.
The following values of the pseudoscalar, vector and tensor mixing angles
have been obtained: $\vert\varphi_P\vert$ = 42.3$^{\circ}$$\pm$3.5$^{\circ}$, 
$\vert\varphi_V\vert$ = 10$^{\circ}$$\pm$8$^{\circ}$ and
$\vert\varphi_T\vert$ = 16$^{\circ}$$\pm$11$^{\circ}$,
in good agreement with the present experimental evidence. 
Also two values of the scalar mixing angle have been obtained:
$\vert\varphi_S\vert$ = 13$^{\circ}$$\pm$9$^{\circ}$ 
if the $f_0$(980) is taken with the real mass, 
but $\vert\varphi_S^{\,bare}\vert$ = 
73$^{\circ}$$\pm$7$^{\circ}$$\pm$24$^{\circ}$
if the $f_0$(980) is taken with the ``bare mass'' of the 
$f_0^{bare}$(720$\pm$100) state.
Only the second value is consistent with recent phenomenological analyses. 
{\it If their conclusions are correct}, it means that in the framework
of our approach the total production rates of scalar mesons are probably
given by the ``bare masses''.

\end{document}